\documentclass[twocolumn,aps,superscriptaddress,showpacs,floatfix]{revtex4-2}
\usepackage{amsmath}
\usepackage{amssymb}
\usepackage{amsthm}
\usepackage{booktabs}
\usepackage{graphicx}
\usepackage{bm}
\usepackage[colorlinks=true,linkcolor=blue,citecolor=blue,urlcolor=blue]{hyperref}

\newtheorem{proposition}{Proposition}

\begin{document}

\title{A Four-Section Bracket for the 48-team World Cup}

\author{Chong Qi}
\email{chongq@kth.se}
\affiliation{Department of Physics, KTH Royal Institute of Technology, SE-106 91 Stockholm, Sweden}

\date{\today}

\begin{abstract}
The expansion of the FIFA World Cup to 48 teams in 2026 introduces structural challenges in tournament design. To populate a 32-team knockout bracket from 12 groups of four, the current FIFA rules select the eight best third-placed teams using a global ranking across all groups. This global coupling creates several major problems: a combinatorial explosion of 495 possible bracket configurations; a fundamentally biased and unequal selection of third-placed qualifiers; lack of a clear path for group winners; vulnerability to collusion and ranking manipulation; and no guarantee of same-group separation beyond the first knockout round. We propose a simple unified solution called the four-section bracket (FSB) rule: split the 12 groups into four sections of three groups. All group winners, runners-up, and the two best third-placed teams in each section advance. Group winners remain in their home sections as local anchors, while lower-ranked qualifiers are transferred to other sections according to a fixed, symmetric rule. This structure guarantees same-group separation until the semifinal, protects the top eight group winners with a predictable knockout path, and reduces bracket complexity from 495 configurations to just one invariant topology per section, recovering the symmetry of the traditional 32-team format. We show substantial improvements in competitive fairness and scheduling predictability.
\end{abstract}

\maketitle

\section{Introduction}

The design of a major international contest should reflect the probabilistic winning chances of its participants, who possess vastly disparate qualities that make the tournament environment inevitably asymmetric. An optimal design must map this asymmetry onto a structured bracket that favors both the highest-ranked teams (the ``supposed best'' prior to the tournament) and the best-performing teams (rewarding outstanding group-stage performance), protecting them from premature elimination while forcing lower-ranked competitors to face a steeper path. In practice, a tournament is a discrete dynamical system subject to competing optimization pressures: competitive fairness, logistical feasibility, resistance to strategic manipulation, entertainment, and commercial viability. Historically, tournament formats based on power-of-two configurations—most notably the 16-team and 32-team World Cups—provided structural symmetry to an asymmetric field of competitors. For three decades, the 32-team FIFA World Cup represented a near-optimal configuration. Dividing $N=32$ teams into $m=8$ groups of four yielded a perfectly balanced binary elimination tree where group winners were rewarded with matches against runners-up, and top-seeded teams were kept apart until the final stages. In this classic design, a progressive path ensured that seeding merit and group-stage performance translated into a predictable knockout advantage.

\begin{figure}[t]
  \centering
  \includegraphics[width=0.95\linewidth]{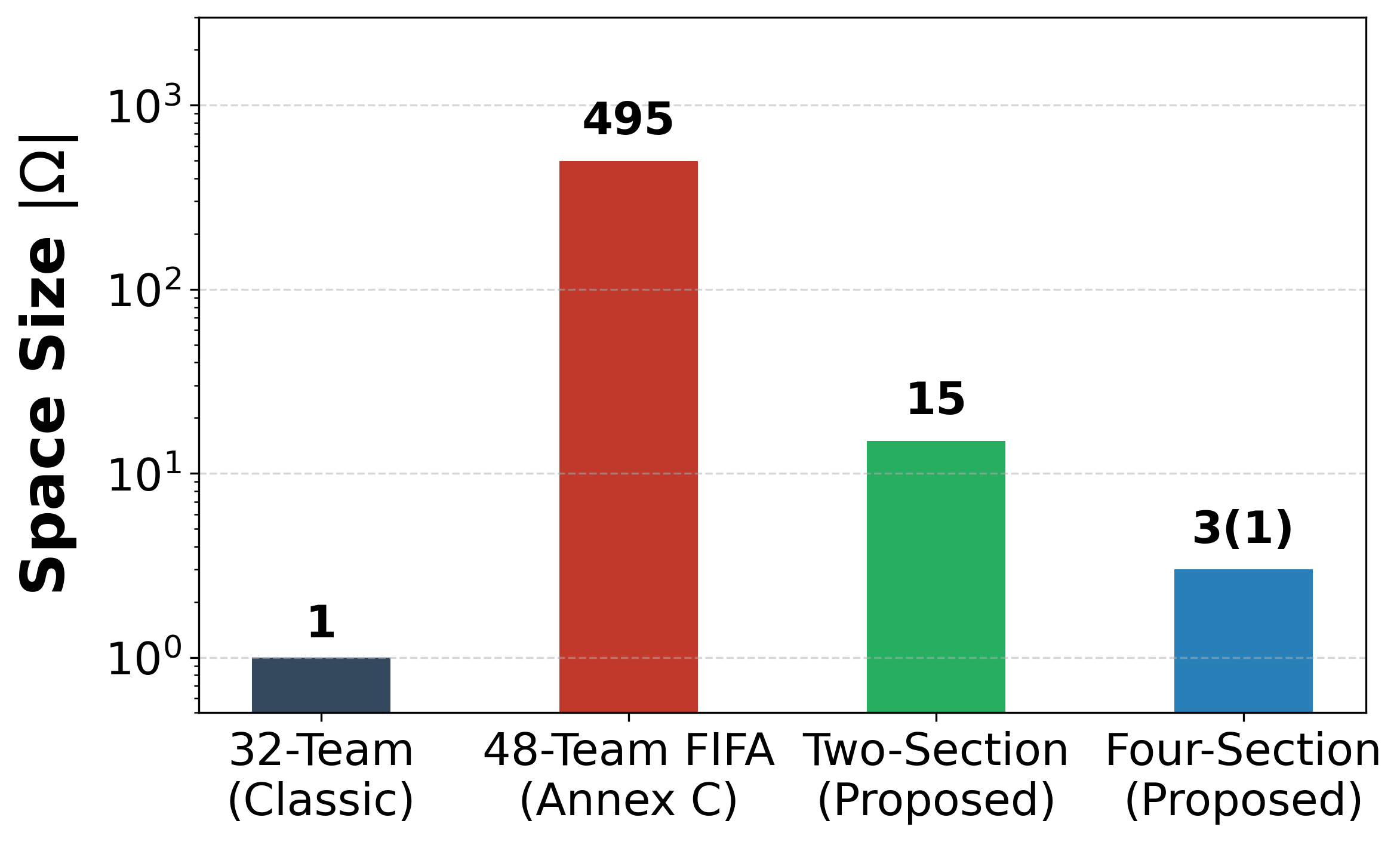}
  \caption{Topological configuration space size $|\Omega|$ for the different architectures. The proposed multisectional designs (Two-Section with $|\Omega|=15$ and Four-Section with $|\Omega|=3(1)$) significantly reduce the combinatorial complexity of the bracket configuration space relative to the globally coupled FIFA system ($|\Omega|=495$).}
  \label{fig:omega}
\end{figure}

The expansion of the UEFA European Championship to 24 teams, and the FIFA World Cup to 48 teams, shatters this structural symmetry~\cite{FIFADraw2026}. Populating the 32-team knockout bracket requires advancing the 12 group winners, the 12 runners-up, and 8 of the 12 third-placed teams. The selection of these 8 third-placed teams is governed by a global ranking across all groups, introducing a non-local coupling with no analogue in power-of-two designs. Furthermore, the draw procedure itself in the 48-team format is subject to complex geographical and seeding constraints, which make the allocation of teams non-uniform~\cite{csato2026nonuniformity,RobertsRosenthal2024,CsatoGyimesi2025}.
Squeezing 12 groups of four teams into a standard 32-team single-elimination knockout bracket introduces severe structural distortions which include:
\begin{itemize}
	\item \textbf{Dilution of group-stage merit:} Since the qualification probability is $66.7\%$ (with up to three teams advancing per group), teams may favor risk minimization over victory~\cite{Csato2025tanking,Stronka2024}. When a single win or even draws are statistically sufficient to advance, the incentive to win is devalued. Teams may prioritize conservative play, defensive preservation, and card or injury avoidance, diluting the competitive intensity of the group stage.
	\item \textbf{Unequal selection of third-placed qualifiers:} Selecting eight third-placed qualifiers via a global cross-group ranking is inherently biased. A third-placed team's advancement depends heavily on the weakness of the fourth-seeded team in its group rather than its own performance. Furthermore, comparing teams across decoupled groups is mathematically problematic, generating 495 distinct combinations of advancing groups (Fig.~\ref{fig:omega}). The resulting combinatorial complexity makes it impossible to pre-determine tournament paths, leaving teams, fans, and host cities with unpredictable schedules and logistics.
	\item \textbf{Lack of a progressive path:} In the Round of 32, eight group winners face third-placed qualifiers, while the remaining four face runners-up. Crucially, the eight winners who face third-placed teams in the Round of 32 are immediately paired against each other in the Round of 16 (Winner vs. Winner). Consequently, half of these top-performing teams are eliminated before the Quarterfinals, failing to establish a progressive advantage for group-stage excellence. 
	\item \textbf{Incentives for strategic underperformance:} By failing to protect group winners past the first knockout round, the bracket can incentivize elite teams to finish as runners-up. The official bracket allows up to four runners-up to reach the Quarterfinals without ever facing a group winner, making ``tanking'' on Matchday 3 a rational strategy to obtain an easier path~\cite{Csato2025tanking,Guyon2020}.
\end{itemize}

To resolve these flaws, we propose the FSB rule, which restores a progressive winning path. The framework divides the 48 teams into four self-contained sections of 12 teams. Each section processes its competitors independently, yielding two quarterfinalists to populate a balanced Round of 8. Crucially, same-group separation is guaranteed via a clean inter-sectional transfer rule. We treat this procedural transfer not as an ad-hoc adjustment, but as a systematic design choice that enhances variety while maintaining mathematical symmetry.

To quantify the topological complexity of the bracket, we analyze the configuration space $\Omega$, defined as the set of possible combinations of qualifying third-placed groups that map to the knockout bracket. As shown in Fig.~\ref{fig:omega}, the globally coupled FIFA system requires selecting 8 out of 12 third-placed teams, yielding $|\Omega_{\mathrm{FIFA}}| = \binom{12}{8} = 495$ combinations. This reduces to 15 under a two-section architecture. Under the FSB rule, the 12 groups are decoupled into four self-contained sections of three groups. Within each section, 2 of its 3 third-placed teams advance, reducing the local configuration space to $\binom{3}{2} = 3$ combinations. Furthermore, because our transfer rules exchange these third-placed qualifiers with other sections, the local bracket topology remains completely invariant of which specific third-placed teams qualify. All third-placed positions in the home bracket are populated by external teams via fixed, symmetric rules. The effective local configuration space size is thus exactly one, denoted as $3(1)$ in Fig.~\ref{fig:omega}.

\label{sec:current}
Under the official 2026 FIFA World Cup Competition Regulations~\cite{FIFA2026rules}, 32 teams advance from the group stage: the winner and runner-up of each of the 12 groups advance automatically (24 teams), alongside the eight best third-placed teams ranked globally by points, goal difference, goals scored, and fair play record~\cite{WikiKnockout2026}.
    
\paragraph{Unequal group-winner treatment.}
Eight group winners face a third-placed team in the R32, while four face a runner-up. A runner-up is structurally a stronger opponent. Moreover, for the eight winners facing a third-placed team, the opponent's identity varies across scenarios, making the match difficulty scenario-dependent and unknowable beforehand.

\paragraph{No same-group separation guarantee.}
The official bracket ensures same-group separation only in the R32~\cite{WikiKnockout2026}. Beyond this round, same-group teams can meet as early as the R16 or Quarterfinals, undermining the ranking established in the group stage.

\paragraph{Variable third-placed opponent clusters.}
The eight open slots can be filled in many combinations, each with different travel, scheduling, and seeding implications. Matchups are finalized only after all 72 group matches finish, using a complex pre-specified lookup table~\cite{WikiKnockout2026}.

\subsection{The cross-group comparison problem}

The core challenge is that third-placed teams are ranked across groups despite playing completely different opponents, introducing structural distortions. We formalize this by modeling the tournament as a mechanism mapping match outcomes to a set of qualifiers.

This cross-group comparison suffers from a fundamental social-choice impossibility. A fair cross-group ranking rule should satisfy three properties: (i)~\emph{independence} (the comparison between any two third-placed teams depends only on results within their respective groups), (ii)~\emph{responsiveness} (performing better in a match always improves, or at least does not harm, a team's advancement probability), and (iii)~\emph{strategy-proofness} (no team has a strategic incentive to underperform, and eliminated teams cannot influence outcomes). These properties are mutually incompatible. If a rule satisfies independence, it must evaluate a third-placed team using statistics from its own group. To be responsive, it must count matches against the fourth seed. However, because the rule is independent, it cannot adjust for the varying strengths of fourth seeds across groups. Consequently, an already-eliminated fourth-placed team—with no competitive incentive—can unilaterally decide the qualification of the third-placed team in its group by conceding more or fewer goals. This moral hazard violates strategy-proofness, showing that no independent and responsive cross-group ranking can be strategy-proof under global coupling.

Under the standard FIFA rule, all three group matches are aggregated. Including matches against the fourth seed (the ``pot-4'' team) in the tiebreakers creates a ``pot-4 donor'' effect. A weak fourth seed acts as a resource donor, allowing the other three teams to accumulate high goal differences. Since fourth-seed strengths are not uniform across groups, the system systematically rewards third-placed teams drawn into groups with exceptionally weak fourth seeds, rather than rewarding competitive merit.

\begin{figure}[!htbp]
  \centering
  \includegraphics[width=0.98\linewidth]{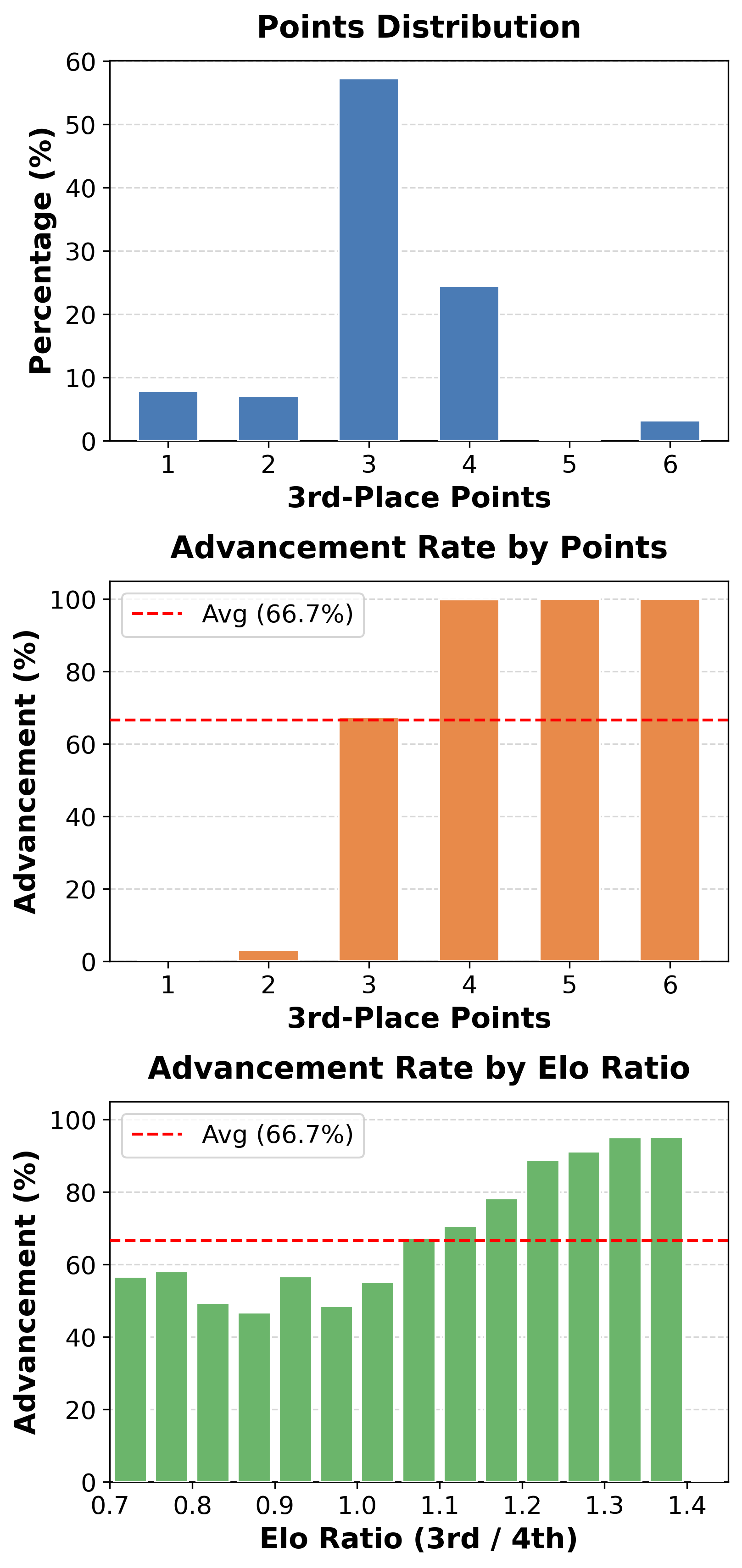}
  \caption{Simulation results under the standard FIFA tiebreaker rule ($200{,}000$ runs). Top: percentage distribution of points accumulated by third-placed teams. Middle: advancement rate of third-placed teams as a function of their points total, showing that $3$ points represents the critical qualification threshold ($67.4\%$ advancement rate). Bottom: advancement rate of third-placed teams as a function of the Elo ratio between the third- and fourth-placed teams in their group. The upward trend illustrates the ``pot-4 donor'' effect, where groups with relatively weaker fourth seeds (higher Elo ratio) yield significantly higher advancement rates.}
  \label{fig:simulation_results}
\end{figure}

To quantify these issues, we simulated the 2026 group stage $200{,}000$ times using a match outcome model calibrated with pre-tournament Elo ratings from June 12, 2026~\cite{EloRatings}. Match probabilities are determined via the ordered logit model of Hvattum and Arntzen~\cite{hvattum2010}, with Poisson goal-generation to resolve tiebreakers. The results are presented in Fig.~\ref{fig:simulation_results}.

Our simulations show that third-placed points are centered around 3 points, which acts as the critical qualification boundary (Fig.~\ref{fig:simulation_results}, top/middle). A third-placed team with 3 points has a $67.4\%$ chance of advancing, whereas teams with 4 or more points advance with near certainty.

Crucially, the results confirm the paradox: a team's advancement probability depends heavily on the weakness of its group's fourth seed. As shown in Fig.~\ref{fig:simulation_results} (bottom), the advancement rate increases dramatically with the Elo ratio between the third and fourth seeds in the same group. This demonstrates the ``pot-4 donor'' problem, where groups with weak fourth seeds act as resource pools that artificially inflate the goal differences of third-placed teams, rewarding draw luck over merit.

\subsection{Robustness to alternative ranking criteria}

To test robustness, we simulated a system where tiebreaker statistics only aggregate matches played against the top two teams in each group, excluding the fourth-seeded team. While this reduces the direct pot-4 donor effect, the overall qualification probability still depends heavily on match outcomes and goals scored in other groups. Because the qualification threshold is determined by a global comparison across decoupled groups, the mathematical dependency on other groups remains. Altering the ranking criteria does not resolve the violation of independence; it merely shifts the bias from the fourth seed's weakness to the competitive parity of groups elsewhere.

\subsection{Asymmetric information and scheduling distortions}

A practical scheduling constraint is that all group-stage matches cannot be played simultaneously. Under the 48-team format, matches must be played sequentially over 10 to 12 days. Consequently, teams playing in the final matches—particularly the eight teams in the last four groups—possess a significant informational advantage~\cite{GuajardoKrumer2023}. Having complete knowledge of the results and goal differences from earlier groups creates asymmetric information. This allows late-playing teams to strategically adjust their play (e.g., settling for a specific draw) to guarantee qualification or select a preferred bracket path. This structural asymmetry violates competitive fairness and is unsolvable under global coupling.
\section{The proposed architecture: four-section bracket}
\label{sec:proposal}

In the official FIFA design, the allocation of the eight best third-placed qualifiers to their R32 slots is governed by a global lookup table. While group winners and runners-up occupy fixed slots, these slots were not designed to prevent same-group rematches beyond the R32. Furthermore, the pairings in the R32 and R16 do not follow a consistent, seed-based protection principle: several top-seeded Pot~1 teams are set on an unavoidable collision course in the R16, while some runners-up face paths that completely avoid top-ranked group winners.

In the FSB design, the 12 groups are partitioned into four sections of three groups each. Third-placed qualifiers are selected locally within each section (2 out of 3), not globally. This single structural change simultaneously resolves the combinatorial complexity, the same-group rematch problem, and the seeding asymmetry.

The FSB architecture differs from the FIFA design in only one core respect: the scope of third-place selection (local section vs. global tournament) and the resulting bracket positions of runners-up. The designer can either keep sections isolated until the semifinal or implement inter-sectional transfers to guarantee same-group separation. In the latter case, the three group winners remain in the home section (with two acting as protected anchors), while the remaining automatic and third-placed qualifiers are transferred to other sections according to a fixed, symmetric cyclic rule.

\subsection{Section structure and pairing}

The 12 groups are organized into four sections, S1, S2, S3, and S4. To ensure competitive balance across the tournament, a natural design choice is to identify a fixed number of top-ranked teams (e.g., 4 or 8 ``super seeds'') based on a pre-tournament ranking of all 48 teams (from 1 to 48). The top four teams (1--4) are distributed as the primary super seeds across Sections S1--S4 respectively, followed by a seesaw or rotational distribution for the subsequent seeds. Under this scheme, one can choose to place seeds 1 and 5 in Section S1, or seeds 1 and 8 in Section S1, as a matter of design and scheduling preference.

Specifically, the assignment of the 48 ranked teams to the sections and groups can be governed by two distinct competitive balance principles:
\begin{enumerate}
    \item \textbf{Cyclic distribution}: Under this principle, the 48 teams are allocated cyclically to the four sections based on their ranks modulo 4. Section S1 receives teams ranked $1 \pmod 4$ (ranks 1, 5, 9, \dots, 45). Within S1, these are distributed to groups $X$, $Y$, and $Z$ such that Group $X$ contains ranks $\{1, 13, 25, 37\}$, Group $Y$ contains $\{5, 17, 29, 41\}$, and Group $Z$ contains $\{9, 21, 33, 45\}$.  The motivation behind this cyclic distribution is to guarantee that every section contains exactly one team from each of the 12 strength tiers (each tier containing four teams), creating an extremely uniform strength profile across all four sections and avoiding any ``group of death'' configurations.
	\item \textbf{Seesaw distribution (snake method)}: Alternatively, if the primary goal is to balance the strength of the top contenders within each section, the designer can pair the top 8 seeds such that the sum of the top two ranks in each section is constant (equal to 9). This leads to the pairings: S1 receives $\{1, 8\}$, S2 receives $\{2, 7\}$, S3 receives $\{3, 6\}$, and S4 receives $\{4, 5\}$. The remaining ranks (9 to 48) are then distributed to balance the overall group strengths. The motivation for this sum-balancing principle is classical competitive fairness: the highest-ranked team (1) is rewarded by facing a lower-ranked super seed (8) in its section, while the most closely matched super seeds (4 and 5) are placed together, establishing a fair gradient of difficulty that mimics traditional binary bracket seeding.
\end{enumerate}

For simplicity and to highlight the mathematical symmetry, we denote the three groups in any given section using the generic symbols $X$, $Y$, and $Z$, rather than specific group letters A to L. In each section, exactly 8 teams advance to populate a local single-elimination bracket consisting of 4 matches (equivalent to the Round of 16 for that section):
\begin{itemize}
    \item The three group winners: $x_1$, $y_1$, and $z_1$.
    \item The three group runners-up: $x_2$, $y_2$, and $z_2$.
    \item The two best third-placed qualifiers from the section, denoted generically as $w_1$ and $w_2$.
\end{itemize}

Under this structure, if we assume that $x_1$ and $y_1$ correspond to the two super seeds in the section, a standard seeding protection principle dictates that they should meet the two qualified third-placed teams ($w_1$ and $w_2$). In practice, because all advancing third-placed teams are qualified based on a common comparison, it does not matter which specific groups these third-placed teams originally came from. 

\begin{figure}[t]
  \centering
  \includegraphics[width=0.99\linewidth]{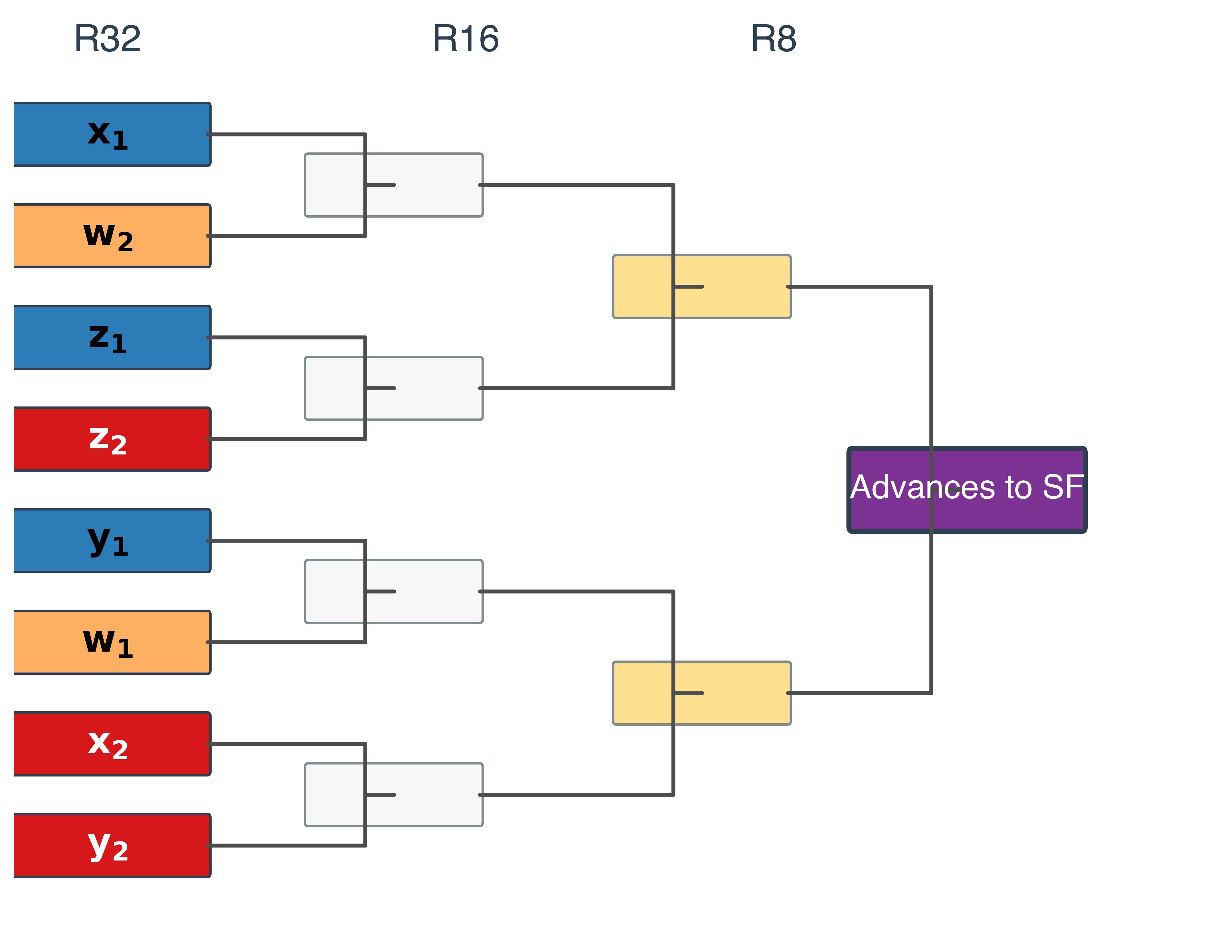}
  \caption{Local knockout bracket for a generic section. The three group winners $x_1$, $y_1$, $z_1$ remain in their home section, while all runners-up and third-placed qualifiers are interchanged with other sections to prevent early same-group rematches. The two super seeds $x_1$ and $y_1$ are protected to face third-placed qualifiers $w_2'$ and $w_1'$; the remaining four slots are filled by $z_1$ and transferred runners-up $x_2', y_2', z_2'.$ Regardless of the transfer rule, this structure provides a clear path for the group winners to advance to the quarterfinals and semifinals.}
  \label{fig:structure}
\end{figure}

For the remaining matchups in the local 8-team bracket, the designer must pair the remaining four slots: the third group winner $z_1$, and the three runners-up (which, after inter-sectional transfer, are denoted generically as $x_2', y_2', z_2'$). The pairing of these teams is a matter of design choice. For example, one can pair $z_1$ against $z_2'$, and have $x_2'$ and $y_2'$ face each other (as shown in Fig.~\ref{fig:structure}); alternatively, one can pair $z_1$ against $x_2'$, and $y_2'$ against $z_2'$. Such permutations do not affect the underlying symmetry or qualification probabilities, and can be adjusted to satisfy regional variety or rematch restrictions.

\begin{proposition}
	No tournament design can protect all 12 group winners equally once
	the field contains a non-power-of-two number of teams.
\end{proposition}

In the FSB design, the eight home anchors (2 per section: $x_1$ and $y_1$ in each of S1--S4) always face a third-placed team in the R32. In the FIFA design, four of the 12 group winners are pre-assigned to face a runner-up in the R32, which is structurally a harder match. The remaining eight FIFA group winners face a third-placed team, but the identity of that team varies across scenarios. Half of these eight winners are eliminated at the R16, when they are forced into winner-versus-winner matchups.
The FSB design improves on this in two respects:
\begin{itemize}
    \item \textbf{Full protection for eight winners:} The eight home anchors always face a third-placed opponent in the R32, with no scenario dependence. These slots are reserved for the tournament's top seeds, giving the strongest teams a predictable and favorable opening path.
    \item \textbf{Transparent, fixed difficulty for the remaining four:} One group winner per section faces a runner-up in the R32. While this is a harder matchup, the asymmetry is fixed, transparent, and known from the moment of the draw.
\end{itemize}

\subsection{The transfer rule and same-group separation guarantee}
The decoupling of third-place selection into independent sections reduces the configuration space from 495 to 3. This reduces to a single invariant configuration once transfer rules are applied to prevent early same-group rematches. Under our framework, same-group separation is guaranteed by a simple, pre-decidable transfer rule:
\begin{enumerate}
  \item \textbf{Winners $x_1$, $y_1$, $z_1$}: remain in the home section to anchor the local bracket.
  \item \textbf{Runners-up $x_2$, $y_2$, $z_2$}: are transferred to the opposite-side sections (which lie in the opposite half of the global bracket). Consequently, a group winner and the runner-up from the same group can meet only in the tournament final.
  \item \textbf{Third-placed qualifiers $w_1$, $w_2$}: are transferred to the SF-paired section (same semifinal half, opposite section). Consequently, a group winner and the third-placed team from the same group can meet at the earliest in the semifinal.
\end{enumerate}

\begin{proposition}
Under the transfer rule above, no two teams from the same group
can meet before the semifinal.
\end{proposition}

The transfer rule provides systematic same-group separation, a property entirely absent in the official FIFA design. The bound is tight: a group winner and the third-placed team from the same group can meet in the semifinal if both advance, while the winner and runner-up from the same group can meet only in the final.

\subsection{Simulation of the proposed architecture}
\label{sec:sim_proposal}

To quantify the competitive impact of the FSB architecture, we extend the Monte Carlo framework to this format. The group-stage simulation (using the same Elo ratings and match model) is identical: 12 groups of four play a round-robin, and the qualifiers are identified. The knockout stage is then resolved using the FSB rule, with the local brackets and transfers implemented.

We analyze the tournament-winning probability for teams at different seeding levels. The results confirm that protecting the two super seeds in each section translates into a meaningful competitive advantage. Under the FIFA design, the four group winners forced to face runners-up in the R32 have their winning probabilities substantially reduced, and the remaining eight are eliminated at a high rate in the R16 due to winner-versus-winner matchups.

In the FSB design, the two protected super seeds in each section ($x_1$, $y_1$) face a third-placed qualifier in the R32, followed by a local quarterfinal against a runner-up or the third group winner. This yields a tournament-winning probability distribution that closely aligns with pre-tournament Elo strength, shifting the winning probability of the top four seeds by up to 20\% compared to the FIFA design.

\section{Conclusion}
\label{sec:conclusion}

We have analyzed the 48-team FIFA World Cup format and documented several structural flaws in its current design: a combinatorial explosion of 495 possible bracket topologies, a significant seeding bias due to the pot-4 donor effect, and the absence of same-group separation beyond the R32.

The proposed FSB architecture resolves these issues by partitioning the 12 groups into four sections of three, keeping group winners in their home sections, and transferring runners-up and third-placed qualifiers to other sections. This design differs from the official format in only one respect: the scope of third-place selection (local section vs. global tournament). The consequences, however, are substantial: the global lookup table is replaced by a simple three-row local assignment; same-group rematches are prevented before the semifinal; and eight of the 12 group winners are guaranteed a predictable, favorable R32 path.

This framework generalizes to any multi-stage tournament where a non-power-of-two participant count forces a global coupling, and the four-section construction is applicable whenever the participant count allows a partition into sections of equal size.

\end{document}